# QUENCHING OF THE QUANTUM HALL EFFECT IN PbTe WIDE PARABOLIC QUANTUM WELLS

J. OSWALD, G. HEIGL, G. SPAN, A. HOMER, P. GANITZER

*Institute of Physics, University of Leoben, Franz Josef Str. 18*
*A-8700 Leoben, Austria*

D.K. MAUDE, J.C. PORTAL

*High Magnetic Field Laboratory, CNRS, 25 Avenue des Martyrs*
*BP 166 Grenoble, France*

We show that for the case of a many valley host semiconductor an edge channel (EC) related non-local behaviour can persist also in the 3D-regime where the quantum Hall effect (QHE) is already quenched. We demonstrate that the QHE is replaced by conductance fluctuations due to EC backscattering in the contact arms which leads to a fluctuating current redistribution between a dissipative bulk electron system and a less-dissipative EC-system. Both electron systems are located in different valleys of the band structure. The linear increase of $R_{xx}$ with the magnetic field is explained by EC-backscattering in the Hall bar

## 1 Introduction

The formation of quantum Hall (QH) plateaux is well understood by the existence of dissipation less edge channel (EC) transport which is possible because of the absence of back scattering directly in the EC's at the same edge[1]. A back scattering is only possible if there is scattering between opposite edges across the bulk region. This happens if the Fermi level is near the centre of a broadened Landau level (LL) and does not happen if the Fermi level is in a DOS-gap between LL's. Therefore basically the realization of a 2D-system is required for the observation of the QHE. The question if it is possible or not to observe the QHE also in a 3D electron system has been investigated by Störmer et al. who demonstrated that the QHE can be observed in a superlattice (SL) with a well pronounced 3D-behaviour as well[2]. Therefore one can conclude that the 3D-behaviour does not directly destroy the dissipation less EC-transport, but in order to avoid dissipation there must not be any coupling between the EC's at opposite edges. This is only achieved if the Fermi level is in a DOS-gap. Because of the very small subband splitting, a wide (quasi-3D) quantum well is some how similar to the SL-miniband system with the major difference that there are no miniband gaps. Consequently EC-transport should play a role also in the quasi-3D regime of a high mobility wide quantum well (WQW). However, there will be a permanent coupling between opposite EC's and therefore a permanent dissipation must occur.

## 2 PbTe wide quantum wells

The realization of wide quantum wells with a flat potential in the electron channel requires a parabolic bare potential which is difficult to obtain in $GaAs/Al_xGa_{1-x}As$ hetero structures by MBE growth[3]. The use of the host semiconductor PbTe allows to create high mobility WQW's without the necessity of remote doping ($\mu > 10^5 cm^2V^{-1}s^{-1}$) and therefore no technological limit towards true 3D-samples exists. Further details about PbTe can be obtained from reference 4. The most important fact is that we get two sets of subbands and two sets of LL's[5]. For typical sample parameters we get a thickness of the conducting channel of about 330nm and an electron sheet density of about $10^{13}cm^{-2}$. This results in a subband splitting of the order of 1 meV for the small effective mass (2D electron system) and about an order of magnitude smaller for the large effective mass (3D electron system). On this basis we get two possible mechanisms for a quenching of the QHE:

(i) The DOS of the 3D electrons parallels the DOS of the 2D electrons. Even if there would be dissipation-less EC-conduction in the 2D electron system, the transport of the 3D electrons must be dissipative. Therefore the potentials of the contacts will be in general different. This leads obviously to a disappearance of the zeros in $R_{xx}$.

(ii) Since the small subband splitting of the 2D electrons is already of the order of native potential fluctuations, there cannot be well developed DOS-gaps even in the 2D system. Therefore the LL-broadening can be considered as a sort of lateral variation of the Landau energy which results usually in the creation of magnetic bound states (EC-loops) in the bulk region. Therefore a permanent coupling between opposite EC's across coupled EC-loops in the bulk region can be expected[6].

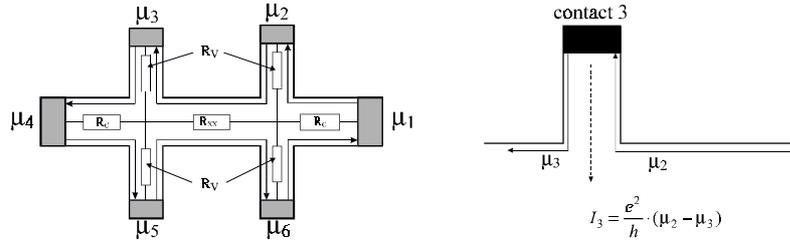

**Figure 1 left:** Equivalent circuit of a Hall bar sample combining EC- and bulk transport. The EC's are represented by arrows connecting directly the reservoirs of the contacts. The bulk system is represented by the resistor network. **Right:** Situation at contact 3. From the difference of potentials $\mu_2$ and $\mu_3$ a net edge current to the reservoir of contact 3 appears. This causes a bulk current (dashed arrow) leaving contact 3.

The situation is schematically explained by the equivalent circuit in Fig.1. Both electron systems are assumed to be independent in the Hall bar and can equilibrate at the metallic contacts only. The necessary suppression of intervalley scattering must be considered as an assumption and cannot be discussed within the scope of this paper. As indicated in the right part of Fig.1, the entering and leaving EC's will be generally at different potential and therefore a net edge current to or from a metallic contacts must occur. This must automatically lead to a compensating current in the 3D electron system. In general we get a current redistribution between the two electron systems at the contacts, which finally determines the total sample behaviour.

### 3. Experiments and discussion

Fig. 2 shows typical data of a magnetoresistance experiment at low temperatures. Further details about the samples and experiments can be obtained from reference 5.

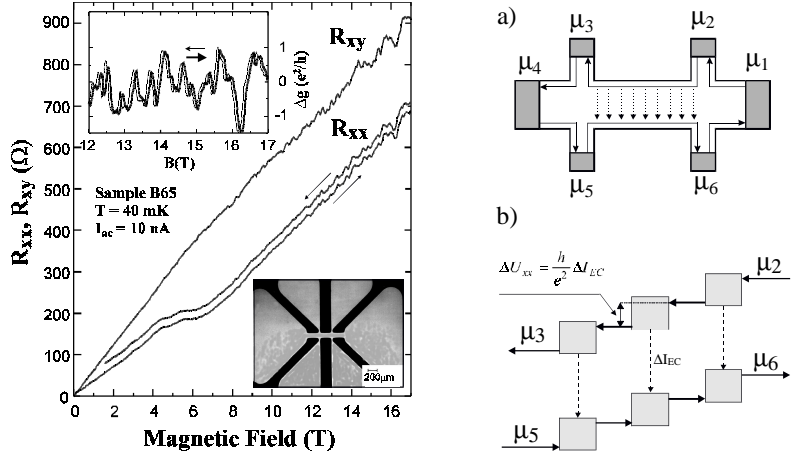

**Figure 2 left:** typical magnetotransport data of $R_{xx}$ and $R_{xy}$. The down sweep of $R_{xx}$ is shifted vertically. Upper insert: Conductance fluctuations extracted from $R_{xx}$. Lower insert: lateral sample structure. **Right:** a) Scheme of a Hall bar sample. b) Potentialscheme in the EC-system due to backscattering. $\Delta I_{EC}$ represents the "edge current" across the bulk region.

The upper field range exhibits perfectly reproducible fluctuations and no indication of zeros in $R_{xx}$ can be found. The corresponding conductance fluctuations (CF) appear to be of the order of $e^2/h$ (upper insert of Fig. 2). These large fluctuations in the macroscopic samples cannot result from diffusive transport in bulk. The amplitude close to $e^2/h$ suggests a contribution of 1D-channels and the most likely candidate are EC's. Additional evidence for the existence of EC's is obtained from recently observed non-local effects in PbTe WPQW`s[7]. However, the QHE seems to be replaced by the fluctuations. The total $R_{xx}$ of the sample is determined by the current distribution between the less dissipative EC-system and the most dissipative bulk system (Fig.1). A current exchange between both electron systems occurs at the contacts because of the net edge current in the contact arms. However, specially in the narrow contact arms there may be also some EC-backscattering which results in a by-passing of the metallic contacts. If e.g. one EC by-passes a contact for some reason, a change of the current redistribution at the contact by it's associated edge current must occur. This results finally also in a change of the bulk current in the Hall bar by the same amount and we get formally a change of a parallel conductivity for $R_{xx}$ by an amount of $e^2/h$. A possible mechanism for the fluctuations is a magnetic field dependent resonant tunnelling across the EC-loops[6] and/or an Aharonov-Bohm (AB) effect due to multiply connected (percolative) paths for scattering of the edge electrons across the bulk region. The effect of a possible EC-backscattering directly in the Hall bar is different and leads basically to dissipation without a current redistribution. This is schematically shown in Fig.2 (right) where EC-backscattering leads to a longitudinal voltage drop also in an EC-system. The increase of $R_{xx}$ with the magnetic field B can be explained in terms of a Landauer-type behaviour: An increase of the Hall voltage with B leads to an increase of $\Delta I_{EC}$ which must finally lead also to a larger $\Delta U_{xx}$. This is basically the behaviour of a QH-system between QH-plateaux and therefore a PbTe WQW is a system which is permanently in a regime between QH-plateaux. A more detailed treatment is beyond the scope of this paper and is considered for publication elsewhere.

**Acknowledgments**

Financial support by Fonds zur Förderung der wissenschaftlichen Forschung (FWF) Austria, Project: P10510 NAW.